# Cooperation and interdependence in global science funding

Lili Miao[1], Vincent Larivière[2,3], Feifei Wang[4], Yong-Yeol Ahn[1], Cassidy R. Sugimoto[5]

[1]Center for Complex Networks and Systems Research, Luddy School of Informatics, Computing, and Engineering, Indiana University Bloomington

[2]École de bibliothéconomie et des sciences de l'information, Université de Montréal

[3]Observatoire des sciences et des technologies, Université du Québec à Montréal

[4]School of Economics and Management, Beijing University of Technology

[5]School of Public Policy, Georgia Institute of Technology, Atlanta, Georgia.

## Abstract

Investments in research and development are key to scientific and economic growth and to the well-being of society[1–3]. Scientific research demands significant resources making national scientific investment a crucial driver of scientific production[4]. As scientific production becomes increasingly multinational, it is critical to study how nations' scientific activities are funded both domestically and internationally[5]. By tracing research grants acknowledged in scholarly publications, our study reveals a shifting duopoly of China and the United States in the global funding landscape, with a contrasting funding pattern; while China has surpassed the United States in publications with acknowledged domestic and international funding, the United States largely maintains its role as the most important global research partner. Our results also highlight the precarity of low- and middle-income countries to global funding disruptions. By revealing the complex interdependence and collaboration between countries in the global scientific enterprise, this work informs future studies investigating the national and global scientific enterprise and how funding leads to both productive cooperation and vulnerable dependencies.

## Introduction

Scientific investments are crucial to national scientific competitiveness[1,6,7]. Cutting-edge scientific research is resource-intensive—requiring facilities, equipment, materials, and labor—making scientific investments a key driver of scientific production[4]. Significant increases in scientific production are often a result of heavy investments in science. For instance, R&D expenditures of China increased at an average rate of 10% per year[8,9] over the last two decades, with total spending increasing from $39 billion in 2000 to $563 billion in 2020[10]. This growth made China the second largest R&D spender at the world level, second only to the United States. Whereas China spent about 11% as much as the United states in 2000, this ratio increased to 84% in 2020[10]. China's investment yielded impressive dividends: while China only accounted for 3.8% of all Web of Science publications two decades ago, it became the largest producer of scientific publications in 2019, surpassing the United States. Although China's publications have long been criticized as having low scientific impact, China also recently exceeded the United States in terms of its numbers of highly-cited publications[11], in part due to its increasing scientific production[12].

In response to the emergence of China, and to strengthen their economic performance and scientific capacity, the EU and the United States have launched massive contemporary science funding programs[13,14]. The CHIPS and Science Act is the latest manifestation of the United States' investment in national science, which explicitly aims to reduce dependency on China for critical technologies[15]. This direct articulation of *dependency* is yet another indicator of the shifting dominance of global science, which moved from Europe to the United States in the twentieth century and is now steadily moving towards China[16–18].

The nationalist rhetoric of scientific competitiveness, however, belies the inherently global nature of scientific production, characterized by the increasing prevalence of international collaboration[19–21]. Scientific articles collaboratively authored by scholars from at least two countries have risen from 14% in 2000[22] to nearly a third of all indexed articles in 2020. These collaborations are uneven across the globe, however: international collaboration constitutes 27% of China's output, 43% of the United States', and 68% of the United Kingdom's. These statistics are not just those of dominance: the highest rate of international collaboration is found in countries with fledgling scientific systems: e.g., international co-authorship in Vanuatu, South Sudan, Liberia, Haiti, and Cambodia exceeded 95% in 2020.

The heavy reliance on international collaboration in many developing countries is attributed, in part, to the lack of domestic funding opportunities[23]. Despite the importance of R&D to scientific development and economic growth, funding for science remains scarce in lower middle-income countries and low-income countries[24]. For instance, in lower middle-income countries (whose GDP value is already much smaller than high-income countries), less than 0.5% of GDP has been used to fund science; the proportion is even lower (at 0.1%) for low-income countries, while the world average is 1.79%[24]. The scarcity of domestic funding is a strong driving force for researchers to seek and rely on international collaboration and foreign funding. National scientific performance, therefore, depends not only on domestic R&D investments, but is also influenced by investments made abroad by other countries[7]. The crucial role of *national* scientific funding and the *global* nature of scientific activities raise an important question: to what extent do nations fund domestic science, and to what degree does each country contribute to global science? What are the countries that underpin the global structure of scientific funding?

Prior research has explored the funding landscape using data on national R&D spending, investigating national scientific performance through R&D spending and the efficiency of turning that investment into knowledge products[2–6,25,26]. R&D expenditures, however, include a wide range of institutions and activities that go beyond basic scientific research, such as applied research and experimental development[27], which accounted for about 73% of R&D expenditures in the United States in 2020[28]. Furthermore, there is no clear agreement on how R&D expenditures should be defined and collected, which hinders a coherent comparison across countries[24]. Most importantly, data on R&D expenditures does not allow for the measurement of how scientific investments flow across international collaboration networks and affect both national and global scientific production and topical profiles.

This paper investigates how countries fund national and international research by tracking research grants disclosed in the acknowledgement sections of scholarly publications. While several scholars examined funding acknowledgement data prior to the inclusion in Web of Science[29,30], it was the advent of indexing of both acknowledgement data and affiliations that made large-scale global analyses possible[31] and led to an increase in such studies[32] with strong implications for science policy[33]. Funding acknowledgement analyses were applied to both localized contexts, such as exploring the concentration of funding in nanotechnology[34,35] and the relationship between funding and innovation in robotics[36], as well as several large-scale analyses examining the relationship between funding and scientific impact[37–40]. Although there are some limitations to these data, validation studies have confirmed the global reliability of the data[35,37,41]. Building upon these studies, we examine publications and funding associated with each country, and quantify how countries support domestic science, cooperate, and rely upon each other for scientific funding as well as countries' vulnerability to shifts and turmoil within the global funding landscape.

## Results

The percentage of publications with funding acknowledgements has steadily increased from 47.7% to 65.1% during the 2009-2018 period (Fig. 1a; see Supplementary Information for robustness analysis). That is, most contemporary articles indexed in Web of Science acknowledge external funding. Given the rise in international collaboration during this same period[42,43], and increased investments in multi-country infrastructures (e.g., the Large Hadron

Collider)[44], one might expect that we would observe a concomitant rise in internationally co-funded articles. This, however, is not the case: only about 10% of publications acknowledge funding from multiple countries and the proportion has remained relatively stable over the last five years (Fig. 1a). The same holds true in internationally coauthored publications: while 73% receive funding, the plurality of internationally coauthored articles (44% of total international collaboration in 2018) report funding from a single country (Fig. 1b). Compared to funding in internationally co-authored research, funding is less likely in domestic science: only about 61% of domestic publications report funding and 57% of domestic publications report funding from a single country in 2018 (Fig. 1c).

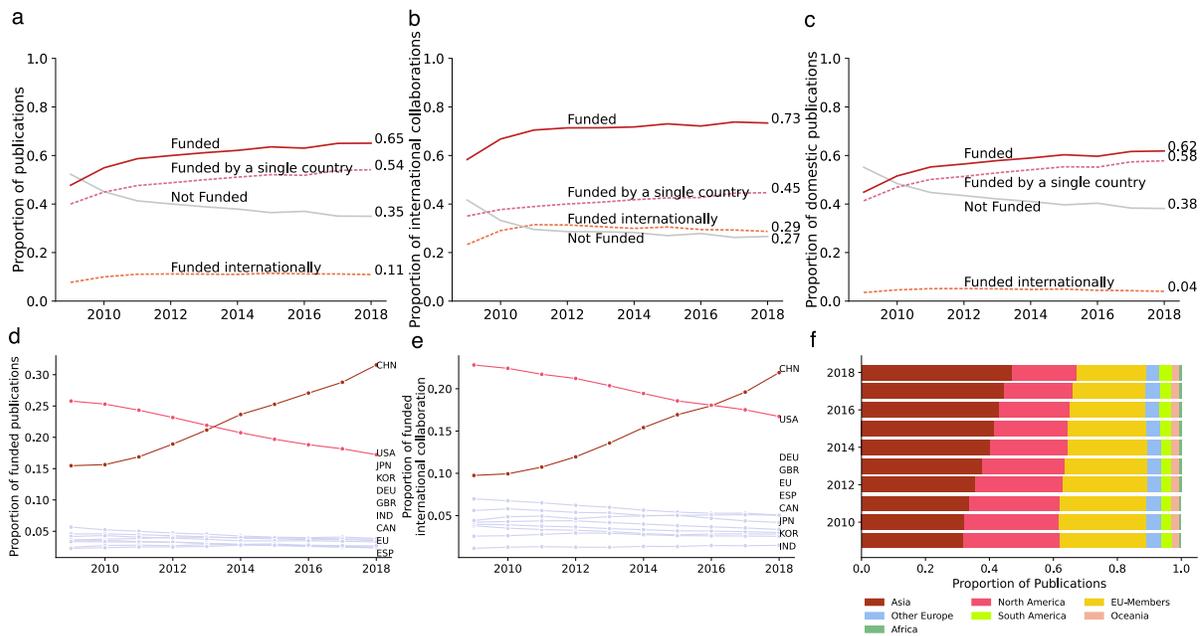

*Figure 1 **Global scientific funding is increasingly dominated by a duopoly structure consisting of China and the United States.** (a) Scientific publications are increasingly funded over the past ten years. By comparing the incidence of papers funded domestically and internationally, we see that most publications are still funded by a single country. The share of publications that are funded by multiple countries remains relatively stable. (b) Same analysis with internationally coauthored publications. (c) Same analysis with domestically authored publications. (d) Proportion of publications that are funded by the top 10 funders. EU refers to the funding organizations that are operated by European Union. (e) Proportion of internationally-coauthored publications that are funded by the top 10 funders. (f) Global share of the funded publications that are contributed by countries across continents from 2009 to 2018 respectively. EU-Members include the funding organizations that are operated by European Union as well as the funding organizations belong to EU-member countries.*

To identify the major funders in research funding at the country level, we measure the proportion of publications that explicitly acknowledge funding from a specific country (see Methods). The results reveal a clear duopoly of the United States and China, with a shifting dominance from the United States to China in recent years. The US was the dominant funder in 2009 when it was acknowledged in 25% of all funded publications, compared with 15% for China (Fig 1d). In 2014, China surpassed the United States and became the largest funder and, by 2018, more than 30% of funded publications acknowledged funding support from China, compared with 17% from the United States. Other Asian and European countries constitute the top ten largest acknowledged funders worldwide (constituting more than 70% of all funded publications); however, they each fund a relatively small percentage of publications and remain firmly behind China and the United States. The observed patterns remain robust to variations in the fidelity of extracting funding data, both temporally and across different countries (see Supplementary Information).

We further investigate the subset of internationally co-authored publications. China has experienced a significant increase in funding internationally co-authored publications and surpassed the United States in 2017 (see Fig. 1e). Asia and North America collectively account for more than 60% of funded internationally co-authored publications, primarily driven by China and the United States. The proportion of funded publications supported by these two countries increased from 41% in 2009 to 49% in 2018. Africa, South America, and Oceania collectively account for about 10% of all funded papers; this percentage is stable throughout the period studied. Overall, the global pattern is characterized by a rapid growth of Asia, a rapid decline of North America, and a slow decline of Europe (see Fig. 1f).

To better understand how each country is supporting its domestic research activities, we define and measure *funding intensity*, which is defined as the proportion of papers, from a country, that explicitly acknowledge funding support (see Methods). Funding intensity varies across countries: for instance, only around 20% of publications in Algeria are associated with funding while the corresponding proportion is 82% in China (see Fig. 2a). However, contrary to the previous research[45,46], we find that that funding intensity across continents remain relatively similar (see Fig. 2b). On average, funding intensity across continents ranges from 53% to 69%, with Asia having the lowest funding intensity and Oceania countries having the highest funding intensity. Scientific publications in the other continents are funded at a comparable level (see Fig. 2b).

We further classified publications based on the author country and funding country to investigate the funding portfolio of countries (see Methods). Although scientific publications in regions such as Africa and Oceania are funded at the similar level of Western countries, domestic institutions fund relatively fewer scientific publications in Africa and Oceania, compared to funding institutions abroad (see Fig. 2c-d; Fig. S4). Of the funded publications, only around 5% of African and Oceania publications are funded exclusively by the authorship countries, which contrasts with the approximately 28% seen in Asia and Europe (see Fig. 2c-d). China stands out as the country with the highest internal funding: among all the funded Chinese publications, 85% of them are exclusively funded by Chinese institutions (see Fig. 2c, Table S1). A similar pattern has been shown in previous articles that publications with Chinese affiliations have higher rate of funding acknowledgement and are associated with higher number of grants[47,48]. In contrast, among all the funded publications that are authored by researchers from the United States, only 63% of them are exclusively funded by US institutions (see Figure 2d, Table S1).

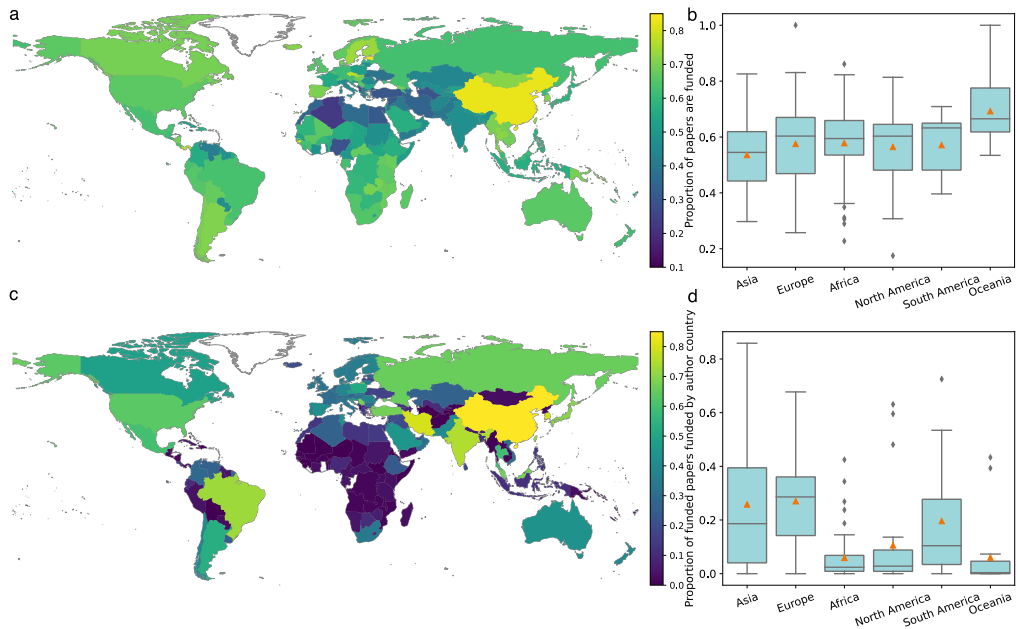

*Figure 2 **Scientific funding intensity across countries. Although, on average, countries across continents have marginal difference in funding intensity, countries differ in terms of the reliance on domestic and external funding.** (a) The funding intensity of countries. To emphasize the variations across countries, the color bar threshold is set at 0.85. Three countries have a funding intensity larger than 0.85. They are Crimea, Niue and Sao Tome & Principe. (b) The distribution of funding intensity of countries across continents. In the box plot, the box is drawn from the first quartile to the third quartile of the distribution. The vertical line represents the median value of the distribution. The lower whisker extends from the box to the smallest non-outlier value within 1.5 times the interquartile range below the first quartile. The upper whisker extends from the box to the largest non-outlier value within 1.5 times the interquartile rang above third quartile. The yellow triangle labels the mean value within each group. (c) The proportion of each country's funded publications that are exclusively funded by the country. China is the only country where around 85% of funded publications are exclusively funded by Chinese funding institutions. (d) The distribution of proportion of funded publications that are exclusively funded by the country itself across continents.*

International collaboration is crucial for creating synergies by combining available equipment, talents, and resources. However, an increased reliance on international collaboration may result in a country depending on foreign resources, and thereby compromising its autonomy. To estimate a country's reliance on foreign funding, we construct various counterfactual scenarios assuming different levels of indispensability for foreign funding in research activities. First, we imagine a simplistic counterfactual scenario where countries are completely cut off from receiving international funding and publications that involve *any* international funding would be eliminated by the withdrawal of international funding, assuming that every acknowledged funding plays a non-negligible role in research activity (see Methods). Under this scenario, we estimate dependence by calculating the proportion of publications that acknowledge at least one international funding instance.

The results show that China and many other Asian countries, as one may expect from their heavy domestic investment, exhibit the least usage of and reliance on international funding. For instance, the proportion of internationally funded publications for China, India, Japan, and South Korea is 11%, 11%, 17%, and 14%, respectively (Fig. 3a). This suggests that the massive investments made by China and India in their domestic science as well as their relative reluctance to internationally collaborate makes them more resilient to changes in international research funding. In contrast, Western countries demonstrate a higher degree of international collaboration and exhibit a more pronounced reliance on international funding (Fig. 3a). For example, 24% of publications by United States and 41% of publications by EU-member countries, on average, would be affected in this counterfactual scenario (Fig. 3b). The corresponding proportion drops slightly to 38% if EU-funding organizations are treated as domestic funding organizations for EU-member countries. Low-income countries, however, are the most dependent on international funding. Despite variations at the country level, we observe that the scientific publications by countries in Africa and Oceania heavily depend on international funding. In these regions, more than half of publications would experience an impact if all international funding were to be removed (Fig. 3b).

However, the assumption that every funding grant plays an indispensable role in research activity overlooks the possibility that additional funding can be leveraged in the absence of others. Therefore, we consider a more stringent counterfactual scenario wherein countries are cut off from receiving foreign funding and only publications *exclusively funded* by foreign sources are influenced. This scenario assumes that only publications that are less likely to leverage the other funding sources would be influenced (see Methods). This scenario does not drastically change the pattern we saw, although European countries show stronger resilience to funding disruption, suggesting that internationally-funded research by European countries tend to be *collaborative*—rather than relying on foreign funding, they tend to draw resources from both the domestic and international sources (see Fig. 3 and Fig. S4-5). By contrast, African and Oceanian countries still exhibit strong reliance, indicating that their current scientific output is much more reliant on international funding (see Fig. S5).

A country's reliance on external funding also means that their research portfolio—*what they publish*—can be largely influenced by the priorities of other countries. A high reliance on external funding may limit the ability of the country to control its own research agenda[5]. As one might expect from the previous results, China and other Asian countries experience the lowest topical profile change (see Fig. 3c-d) in the exclusion of papers with foreign funding. The United States is also among the ten countries least affected by funding removal. A similar pattern holds for many European countries. Although about 40% of publications are linked to international funding for EU-member countries, their research profiles are marginally influenced even if we remove the publications that are internationally funded (see Fig. 3c-d). The most significant influence is observed in Oceanian and some African countries; the topic distribution of research publications produced with international funding is distinct from those that are not associated with international funding. This finding resonates with the concept of "parachute science" in global research, highlighting that the research priorities of developing countries are frequently overlooked in international collaborations with researchers from developed countries[49]. This marginalization is attributed to the power asymmetry in international collaboration, with source of funding serving as a significant factor contributing to this imbalance[5].

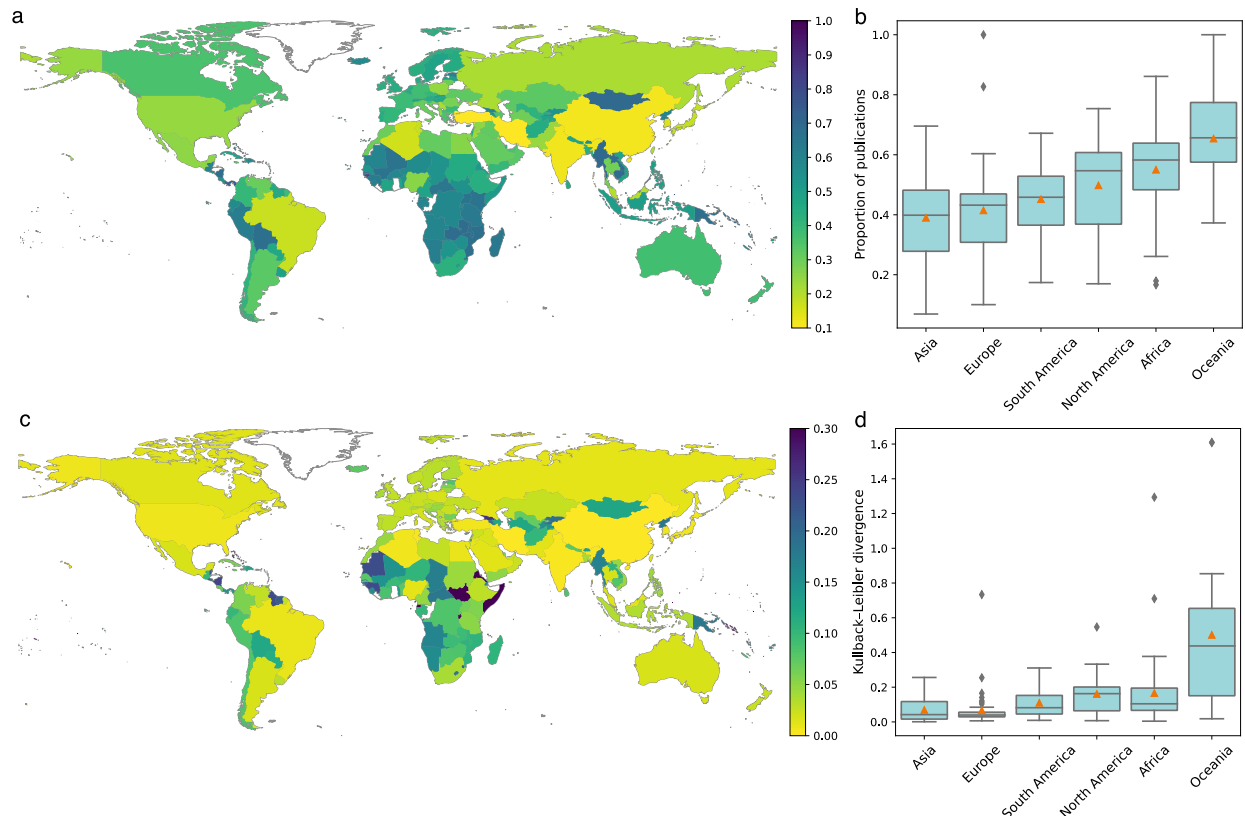

*Figure 3* **The impact of removing internationally funded publications. Asian countries experience the least lost while African countries as well as Oceania countries suffer the largest lost.** *(a) The proportion of publications influenced in each country following the removal of international funding (b) The proportion of publications influenced in each region following the removal of international funding. In the box plot, the box is draw from the first quartile to the third quartile of the distribution. The vertical line represents the median value of the distribution. The lower whisker extends from the box to the smallest non-outlier value within 1.5 times the interquartile range below the first quartile. The upper whisker extends from the box to the largest non-outlier value within 1.5 times the interquartile rang above third quartile. The yellow triangle labels the mean value within each group. (c) The difference between actual research profile and the counterfactual research profile. The difference is measured by the Kullback-Leibler divergence. Large KL-divergence value represents the counterfactual research profile is distant from the actual research profile and vice versa. To highlight the difference among countries, we set the threshold of the maximum value to 0.3 There are 23 countries have KL-value large than 0.3. (d) the profile change of countries by continents.*

We then switch perspective from the "receiver" to "funder" by estimating the impact of each country on others. We measure how much perturbation a specific country can cause to others if it stops participating international funding activity, by using the same two extremes used above (see Methods). First, we count the proportion of publications that have at least one funding acknowledgement from a focal country. The results show that the United States is the most influential country in terms of global funding (see Fig. 4a-b). On average, in 2009, around 12% of publications in each country would be impacted if the United States ceased funding research that involves scientists from other countries. Due to the increasing international collaboration, this figure rose to 17.5% in 2018. EU funding organizations, UK, France, and Germany also have substantial influence over the research activities of other countries. However, the corresponding percentages have consistently remained below 10%. Our results also indicate EU funding organizations play a vital role in the UK's research system: around 10% of British publications would be influenced if EU funding organizations were no longer providing funding to the UK. China, even with its rapid rise in quantity, has limited influence on other countries from this point of view, as other countries would only experience a marginal influence (of slightly more than 5%, on average) if China stopped funding internationally (see Fig. 4a). The influence of countries remains similar when measuring from the ability of altering countries' research profile; countries experience the largest extent of profile change when the United States withdraws from international funding (see Fig. 4b).

Considering the simple counterfactual scenario where a publication would have been affected only if it is exclusively funded by the focal country, we then count the proportion of publications of each country that are exclusively funded by the focal country (Methods). This exercise shows a consistent trend: the United States demonstrates the most substantial impact on the scientific production of other countries, influencing approximately 8% of publications on average in each country. By contrast, the remaining major funders influence less than 3% of publications in each country (Fig. S6).

Yet, our counterfactual scenarios assumes a direct relationship between funding and publications that overlooks the complexity in scientific production. National research production is simultaneously influenced by various factors, including the country's existing scientific capacity[50,51], overall investment[50–52], and the broad scientific environment[50–52]. Moreover, the elasticity of production to domestic or international funding may exhibit a range of possible values. To tackle this gap with our data, we further employ a fixed effects panel regression model to examine the influence of funding from major scientific funders while accounting for other relevant factors (Methods and Fig. 4c). Specifically, we investigate whether the inflow of scientific investment from major funders can predict the scientific growth in countries. The regression results affirm the crucial role of foreign scientific funding in national scientific production, with funding from the United States demonstrating the most significant influence on the growth of scientific production in other countries. As illustrated in model 1 in Figure 4d, foreign scientific funding significantly predicts publication growth rate of countries, surpassing the magnitude associated with domestic funding. This result resonates with our finding that, on average, most countries outside of the existing circle of scientific powerhouses exhibit substantial dependence on external funding (Fig. S4). More specifically, funding from the United States plays a pivotal role, with a 10% rise in the funding from the United States is associated with a 2% increase in the publication growth rate (Fig. 4d). In contrast, funding from China does not significantly predict publication growth of other countries (Fig. 4d).

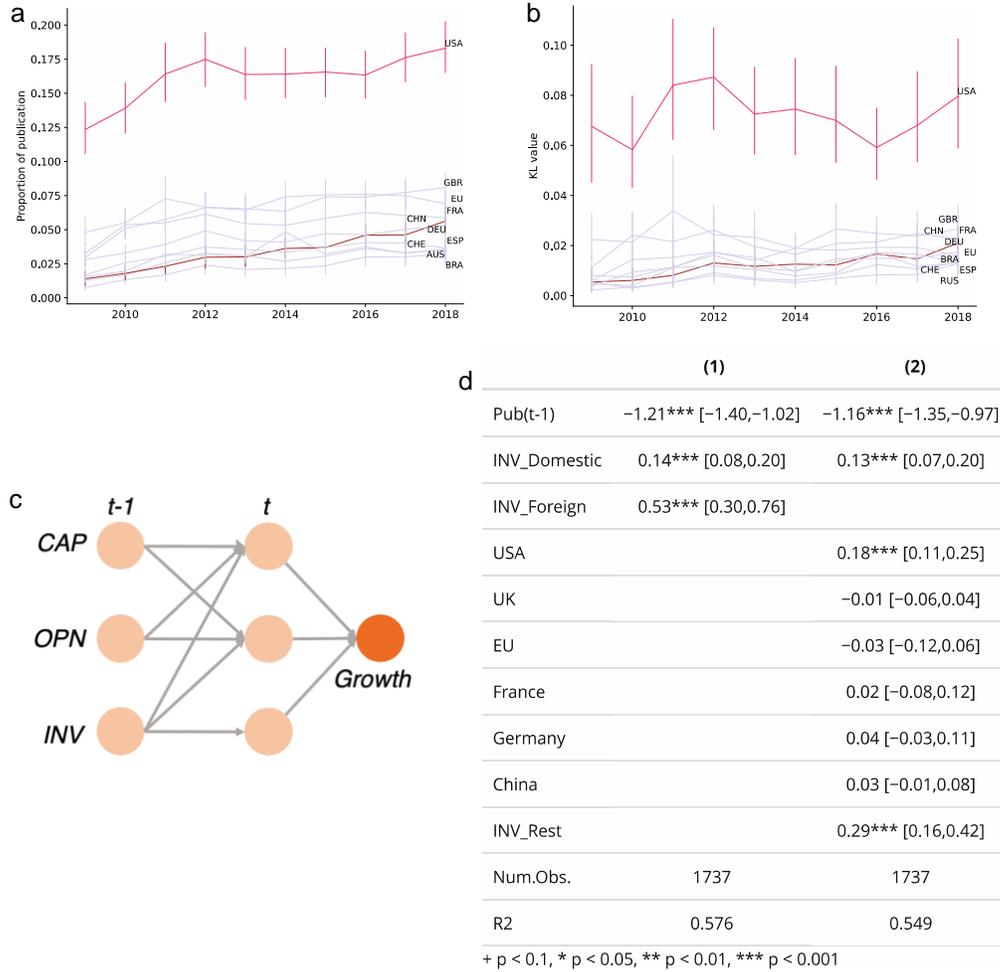

*Figure 4 **The United States has the largest impact on other countries.** (a) The average proportion of publications influenced when internationally funded publications by the focal country are removed. To compare the impact of the United States. with China, the United States. and China are highlighted. The line shows the mean value of each country. Error bars represent the 95% confidence interval of the mean drawn from bootstrapping. (b) The KL-divergence value of research profile of countries when the internationally funded publications by the focal country are removed. (c) The causal diagram on which our regression models are built. Here, $CAP_t$ stands for the scientific capacity of the country at time at time t, $OPN_t$ represents the extend of international scientific cooperation at time t, $INV_t$ represents scientific investment at time t, and $Growth_t$ represents publication growth during time t. (d) Results from the fix-effects regression models.*

To reveal a more nuanced difference of the impact of funding from major funders, we further investigate the *sphere of influence* of the United States, EU, UK, and China. The first three countries and regions are chosen because withdrawing funding from them results in an influence on more than 5% of publications across countries, and China is included for comparison (Methods). The results reveal that removing funding from the United States causes a substantial influence globally, with the most salient influence observed in African countries and

Latin American countries (Fig. 5a). Meanwhile, the United States is considered as the most important funding source by the largest number of countries (Fig. 5a). In contrast, funding organizations from the European Union and from the UK exert influence primarily within Europe, with the impact of UK funding extending to certain African countries and Asian countries with colonial ties, such as India and Malaysia (Fig. 5b-c). Despite China being the largest funder to global science, its impact on global scale remains marginal, and only a select few Asian countries, such as Singapore, Japan, and Vietnam, consider it the most significant source of funding (Fig. 5d).

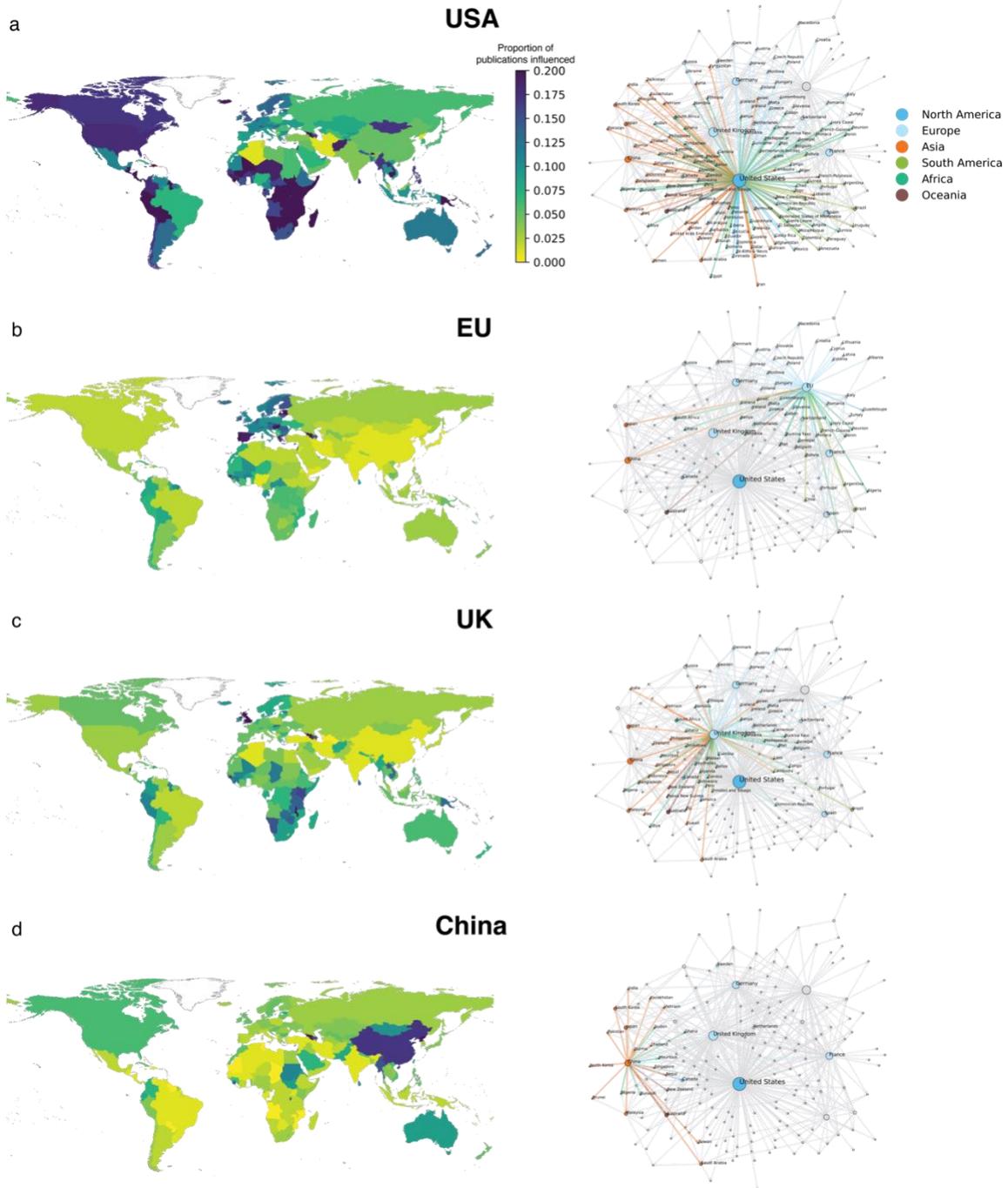

*Figure 5. **Scientific funding from the United States has exerted a significant influence on countries worldwide, whereas China's influence is primarily concentrated in Asian nations.** (a)-(d) Exemplars illustrate the distribution of influence from four major scientific funders. The proportion of influenced publications is calculated as the percentage of publications in each country acknowledging funding from the specified focal funders. The backbone networks illustrate significant funding partnerships between countries, with coloring applied only to countries receiving a substantial portion of their funding from the focal country. Node color corresponds to the six continents.*

## Discussion

National scientific development hinges on the availability of scientific investments[53,54]. However, constrained by limitations and heterogeneity of R&D expenditures data, it has remained challenging to describe the global scientific funding landscape . Using funding acknowledgements disclosed in scientific publications indexed in the Web of Science, our study provides a global-scale analysis of funding structures behind national scientific activity, and interconnections between countries through scientific funding.

We find that the rise of China's scientific system has led to a US-China duopoly in the global scientific funding structure, with a relative decline in the US. Our results reaffirm the observation that researchers in developing countries are under-funded by domestic institutions[5,45], leading to an overreliance on foreign funding. Our analyses suggest that developing countries would lose a large fraction of publications and experience a larger alteration of their research profile if international funding is removed. Even with the rapid rise of China in global stage, the United States maintains the largest influence on the other countries.

Our results demonstrate that nations are deeply embedded in an interconnected global scientific system where they are heavily reliant on each other. Even when controlling for relevant factors, foreign scientific investment continues to demonstrate a significant association with the national publication growth rate. These dependencies, however, are highly asymmetrical, which creates a discrepancy in where science is done and where scientists and investments are from[61], as well as in leadership roles on scientific teams[62]. "Parachute science"[63] or "helicopter research"[64] is the practice whereby scholars, typically from countries with higher scientific capacity, carry out research abroad with little involvement or engagement from the local community. These practices are often the result of colonial relationships, and perpetuate the assumption that rich countries have a right to study and utilize the environment of less resourced nations[65]. To achieve sustainable global development, it is crucial for major scientific nations to recognize their influence on scientific development of other, particularly less-advanced, countries[66].

Our results call attention to the issue of dependence on foreign funding in low-income countries and the potential consequences and threats it poses to future scientific development. Funding underlying global science is linked with the deeper and sustained inequality in global scientific structure[67]. The power asymmetries enforced by scientific funding from high-income countries to developing countries inevitably lead to overlook the research agenda in low-income countries[68]. For example, investment from the US National Institutes of Health in South Africa far exceeds national investment in health research[69], which allows a foreign entity to effectively set the research agenda for the country. Our research reinforces the strong influence of developed nations on the topic space and research profile of developing countries. Therefore, partnerships should also seek to improve capacity building and build joint funding opportunities[70], to lessen asymmetrical global dependencies[71]. To build a productive and sustainable scientific system in developing countries, funders in high-income countries and potential local funders in low-income countries should work collectively to shape a new framework to better fund science.

## Data and Methods

**Publication data**

Publication data is drawn from Clarivate Analytics' Web of Science (WoS) database hosted and managed by the *Observatoire des Sciences et des Technologies* at *Université du Québec à Montréal*. Publications are associated with countries using the institutional addresses listed by authors on their papers. Disciplinary classification of publications is based on the National Science Foundation field and subfield classification of journals, which categorizes each paper published in a given journal into a discipline and a specialty[72]. The classification was further complemented by an in-house classification for the Arts and Humanities[73]. The resulting classification scheme contains 143 specialties, grouped into 14 disciplines: Biology, Biomedical Research, Chemistry, Clinical Medicine, Earth and Space, Engineering and Technology, Mathematics, Physics, Arts, Health, Humanities, Professional Fields, Psychology, and Social Sciences. Considering the incomplete funding coverage in social science and humanities publications[31], we excluded Arts, Heath, Humanities, Professional Fields, Psychology, and Social Sciences from our analysis. We limited our analysis to journal articles and review articles. We also excluded publications that did not contain institutional addresses or disciplinary categories. WoS began indexing funding information during the year 2008; therefore, we began the analysis in 2009. After these filters, the dataset contained 12,759,130 articles published between 2009 and 2018.

**Funding acknowledgement data**

Information on the research funding of a paper was retrieved from the 'Funding Agency' and 'Grant Number' fields in the WoS. We limit our analysis to the funding organization strings that appeared at least twice in the database, given that organizations appearing once are largely spelling mistakes, non-funding organizations[74], or negligible funding agencies. 3,086,974 unique name strings were removed in this step, leaving 756,881 unique name strings. This yielded a reduction of 755,031 articles (i.e., 6% of all articles) from the analysis. The retained strings may include organizations that are acknowledged for contributions other than funding; however, empirical studies suggest these instances are relatively rare[59].

We then used a previously curated dataset and two automatic identification approaches to assign funding organizations to countries. The curated dataset was inherited from a previous study examining the mental health research funding system which includes nationality information of 1,783 (0.2% of total identified institutions) funding agencies[60]. For the remaining institutions, we developed two approaches to automatically identify nationality. First, we used the names of countries and the variations of names within the names of funding organizations. For instance, "China" can be identified from many Chinese funding organizations (e.g., "NSF of China"). Name strings containing "EU" or "European"—such as "European Science Foundation"—are classed as such: considering that EU funding organizations are supported by member countries, we label them as "EU" rather than individual countries. Through this approach, 237,313 (31.4% of total identified institutions) institutions were assigned to a country. 3,764 (0.4% of total identified institutions) name strings contained the name of multiple countries, such as "US-Israel Binational Science Foundation"; these were labeled as "multi-national".

For the remaining strings (59.3% of total identified institutions), we inferred the nationality from the main country affiliation of articles funded by each institution. More specifically, we compiled the distribution of countries found in articles funded by each funder and assigned the country that was most frequent. In most cases, a country appears much more frequent than others. For example, 98% of papers that report the funder string 'NERC' (Natural Environment Research Council) had affiliations from the United Kingdom; the funder was therefore assigned to the UK. Similarly, 98% of papers that report funding from 'UGC' (University Grants Commission) come from institutions affiliated with India; that string was identified as an Indian funding agency. By leveraging authorship institution information, we were able to identify the national affiliation of 438,247 (57.90% of total identified institutions) funding organizations. We exclude 10,453 (1.38% of total identified institutions) organizations from our analysis as they could not be assigned to any single country due to the equal distribution from multiple countries. We applied two approaches to validate the accuracy of our identification (see SI). Although the approach may have a slight bias to assign organizations to more scientifically advanced countries (due to higher production of articles), the validation results show high accuracy of assignment (see SI).

Our final dataset contained 12,759,130 publications; 5,022,190 (39.36% of all publications) publications are not associated with funding information, 6,620,701 (51.89% of all publications) publications are associated with funding organizations that were identified via country name matching, 36,971 (0.3% of all publications) publications receive funding from "multi-national" institutions, 3,644,249 (28.56% of all publications) publications are associated with the institutions that were identified via authorship, and 14,639 (0.11% of all publications) publications are associated with unidentified funding organizations. Since the focus of our study is to understand the source and the destination of the scientific investment across countries, we exclude the "multi-nation" funding institutions and the unidentified institutions from our analysis; those account for 0.41% of the total publications in our analysis. It is important to note that certain types of funders such as government laboratories, charity units and commercial companies are less likely to be explicitly acknowledged by authors. Therefore, in our analysis, "funded papers" refer to those containing explicit funding information, while it's possible that papers without such information may still have been funded.

**Assignation of publications to funding country**

We use fractional counting to assign funded publications to each country, defined as $f_{c,p} = \frac{N_{c,p}}{N_p}$ where $f_{c,p}$ is the proportion of paper $p$ that is funded by country $c$, $N_{c,p}$ is the number of funding instances that come from country $c$, and $N_p$ is the total amount of funding instances that are acknowledged in paper $p$. A funding instance refers to the 'funding agency-grant number' combination recorded in the dataset: e.g., NSF-1904280 and NSF-2144216 are considered as two different funding instances[74]. For the funding agencies without grant numbers, we assume one grant comes from that agency. This conceptually makes each funding instance equivalent, which is a major caveat of this study. However, we note that it is challenging to find a better and feasible alternative. First, it is impossible to identify the amount of every grant consistently and accurately across all countries, no global datasets of funding amounts exist. Second, even if the total funded amount of each grant could be revealed, the amount of direct research funding varies substantially across institutions and countries due to indirect cost. Third, the funding required for a research project can vary greatly across disciplines and countries due to differences in the nature of the involved costs, as well as variations in the costs of labor and materials needed. Finally, the fact that large grants tend to produce more papers partly mitigate the bias from focusing on the funding instances. Given the constraints of available datasets, therefore, we employ acknowledged funding grants as a proxy of countries' funding activity.

**Measuring a country's share of funded publications**

To estimate a country's contribution to global scientific funding, we measure the proportion of global publications that are funded by each country. The proportion of global publications that are funded by a country is defined as $F_c = \frac{\sum_p f_{c,p}}{F}$ where $\sum_p f_{c,p}$ is the sum of the proportion of the funded publications by country $c$ and $F$ is the total number of funded publications globally.

**Measuring a country's research funding intensity**

To investigate the funding portfolio of countries, publications are classified into four groups based on the involved funders after they are assigned to the authorship countries, namely, no-fund-inf, domestic, co-funded, and foreign (Fig. 6). For ease of interpretation, we use the full counting method to assign publications to countries based on authorship[56]. "No-Fund-Inf" refers to publications without any funding information in WoS database. "Domestic" refers to papers that are funded exclusively by the focal author's country. For instance, if a publication has authors from both China and the US, but is funded solely by China, then the publication is viewed as "domestic" funded from China's perspective, whereas it will be classified as "foreign" funded from the perspective of the US, as we will explain shortly. "Co-funded" means the author country participated in the funding activity with other countries, e.g., for a collaborative publication authored and funded by both China and the US, the paper is classified as co-funded for both countries. "Foreign" means the author's country is not listed as the funding country. For instance, for an EU-funded collaborative publication authored by China and the US, the paper is classified as foreign-funded for both China and the US.

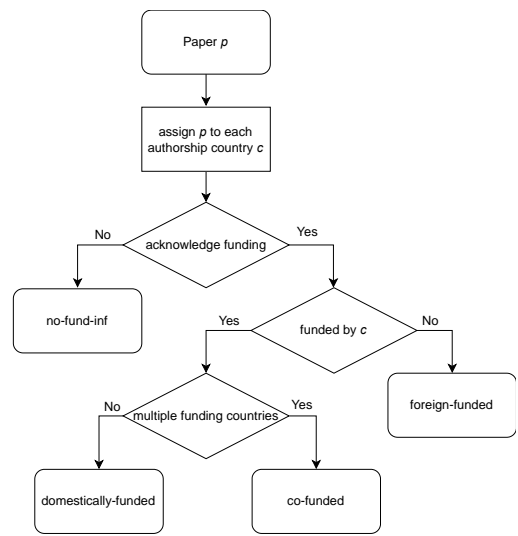

*Figure 6 Classifying publications into four funding types based on the countries providing funding and the countries of authorship.*

The overall funding intensity of a country is defined as $I_c = \frac{1}{M_c}\sum_{m \in M_c} \delta(m, F)$ where $M_c$ is the number of publications that are authored by country $c$, $\delta(m, F)$ is 1 if paper $m$ acknowledges funding regardless where the funding comes from otherwise the value is 0. To characterize a country's gross funding capacity, we measure the proportion of publications that are exclusively funded by the country itself which is defined as $C_c = \frac{1}{M_c}\sum_{m \in M_c} \delta(c, m)$ where $M_c$ is the number of publications that are authored by country $c$, $\delta(c, m)$ is 1 if paper $m$ acknowledges funding solely from country $c$ otherwise the value is 0.

**Estimating a country's dependence on international funding and its global impact**

To investigate a country's dependence on international research funding, for each country, we calculate the percentage of publications that would be influenced if we excluded all internationally-funded publications. Internationally-funded publication refers to any publication that acknowledges funding resources from a country that is different from the focal authorship country (Fig. 7a). For instance, paper $p$—co-authored by China and the United States while funded by China—is considered as an internationally-funded publication for the United States and as a non-internationally-funded publication for the China since Chinese' funding resources flows to US authors who participated in research through paper $p$. Removing internationally-funded publications for a country can be considered as an extreme hypothetical scenario where the country is cut off from receiving funding resource from all foreign countries, influencing publications involving any degree of international funding. We call the publication record without internationally-funded papers as the counterfactual publication record. Meanwhile, considering the potential situation wherein researchers can leverage domestic funding in the absence of foreign financial support, we build the second counterfactual publication record by removing publications exclusively funded by foreign sources (Fig. 7b). This additional experiment estimates countries' dependence on international funding by assuming that only papers exclusively funded by foreign funding would be affected when the country is disconnected from foreign funding.

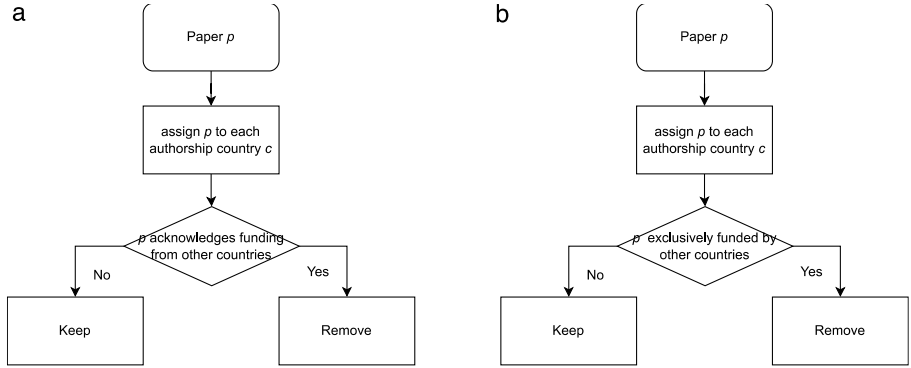

*Figure 7 Illustration explaining the process of determining whether a publication is included in the counterfactual record.*

In addition to examining the number of remaining publications, we also investigate how countries' research profiles are changed by removing these publications. A country's research profile is measured as the distribution of number of publications in each discipline. To estimate countries' dependency on the papers that receive foreign funding, we use the Kullback-Leibler divergence (KL-divergence) between the actual research profile and the research profile after removing internationally-funded publications (counterfactual research profile). The KL-divergence is defined as:

$$D_c(P||Q) = \sum_{x \in \chi} p(x) \ln \frac{p(x)}{q(x)},$$

Where $Q$ is the actual research profile of country $c$, $q(x)$ is the proportion of publications in discipline $x$ in country $c$, $P$ is the counterfactual research profile of country c, $p(x)$ is the proportion of publication in discipline $x$ in the counterfactual profile. $D_c(P||Q)$ measures the extra number of bits required to represent the counterfactual research profile using the code that is optimized for the actual research profile. Large KL-divergence value means the counterfactual profile is more distant from the actual research profile, indicating the topical distribution of internationally-funded publications is more distinct from that of domestically-funded publications. In other words, large divergence suggests that the country's research focus may be largely swayed by foreign funding agencies' priorities.

In addition to measure the general impact of internationally funded publication, we replicate the same analysis by removing internationally funded publications that are funded by a single country, to estimate the impact of a specific country. For instance, to estimate the global impact of funding from the United States, we remove publications that have non-US authors where the US's funding agencies are acknowledged, considering the case where the United States had stopped international funding and the publication would be influenced. After filtering out those publications, we measure the proportion of publications influenced and changes in research profile across countries. Moreover, to address the possibility that researchers can potentially access funding resources from alternate countries in the absence of financial support from a specific country, we build the second counterfactual publication record by removing publications exclusively funded by the focal country. The analysis measures the impact of a country under the condition that the focal country is the sole provider of financial resources necessary for the paper's production.

To understand a country's reliance on funding resources from other nations, we construct a funding reliance network where nodes are countries and directed, weighted edges capture the reliance of one country on the other. For instance, a direct edge from country $c_1$ to country $c_2$ with a weight of 0.2 represents 20% of publications would be influenced in country $c_2$ if country $c_1$ stops funding internationally and if all the publications that was funded by $c_1$ could not be realized. To identify the most influential funders for each country, we apply the multiscale backbone extraction method[75], which uses a simple null-model to identify the most disproportionally significant edges around each node.

**Fixed-effect regression model**

Empirical studies have demonstrated that national innovation capacity is intricately characterized by a nuanced set of observable factors, encompassing inputs devoted into innovation system such as scientific manpower and scientific investment[50–52]. Additionally, the environment for innovative production, such as the extent of IP protection[50–52] and openness to global cooperation[50–52], along with a country's knowledge stock[50,51], play determining roles. Building upon these empirical evidences, our conceptual model posits that national scientific publication growth is a function of scientific production capacity, openness to international cooperation, and scientific investment (Fig. 4c).

Scientific capacity in our model considers available scientific personnel, infrastructure, the stock of accumulated knowledge, and the capability to convert scientific capital into publications. Given the infrequency of abrupt changes in a country's scientific capacity between two consecutive years and recognizing the causal relationship from scientific capacity to the number of produced publications, we approximate scientific capacity at time $t$ with number of publications produced in time $t-1$. Furthermore, we posit that a country's scientific openness is closely linked to the extent of international collaboration and consequent external funding. Consequently, we substitute the openness factor in the theoretical model with the amount of foreign funding instances acknowledged by the papers published by the country. To assess the impact of funding from different countries, we categorize funding sources, namely: China, EU, France, Germany, United Kingdom, United States, and others. The scientific investment of a country is measured by the number of domestic funding instances acknowledged in publications. The fixed-effect model is defined as follows:

$$G_{i,t} = \beta_0 + \beta_1 P_{i,t-1} + \sum \beta_q F_{q,t} + \alpha_t + \alpha_i + \varepsilon_{i,t}$$

Where $i$ denotes countries, $t$ denotes time periods, $G_{i,t}$ is the publication growth rate in the receiving country between time $t$ and $t-1$, $P_{i,t-1}$ is the number of publications produced by country $i$ at time $t-1$, $F_{q,t}$ is the amount of funding instances from each distinctive country include country $c$ itself, $\alpha_t$ and $\alpha_i$ are the time-specific and country-specific intercepts that capture the heterogeneity across time periods and countries.

**Limitations**

This study has several limitations. First, estimations derived from Web of Science are subject to biases and errors. Web of Science, being developed and maintained by the Western, Anglophone scientific enterprise, tends to overestimates research and related funding from Western countries and publications in English while underestimating the production and funding in other nations and languages[56]. Moreover, the effectiveness of the funding indexation algorithm varies across countries and years, leading to underestimations of funding data for earlier years and certain countries (see SI). Nevertheless, given the comprehensive coverage of funding data within the dataset used for our analysis, our results remain largely unaffected by the omission of some information (see SI). Furthermore, the database primarily focuses on journal articles, neglecting alternative forms of output like patents and book projects, which could potentially result in an underestimation of funding-related output.

Second, our analysis is based on the identification of funding acknowledgements within publications. It is important to note that funding acknowledgement information, while a valuable resource, may not comprehensively reflect the entire financial support for knowledge production. Funding from certain types of funders, such as hospitals or government agencies, may be less likely to be explicitly acknowledged, a phenomenon denoted as implicit funding[57,58]. However, given the limited systematic understanding on the prevalence, role, and mechanism of implicit funding, we defer the examination of the impact of implicit funding to future research. Therefore, within the scope of our analysis, "funded publications" specifically refer to those that contain explicit and identifiable funding information. In addition, it is possible that institutions are acknowledged in publications due to various incentive reasons. Authors may cite funding that did not directly contribute to the work but were included to demonstrate evidence of labor for grants (see SI on data section). However, as funding agencies and publishers are increasingly strict about the reporting of financial support behind their publications, funding acknowledgements are now more effective in reflecting the financial investment behind publications[59,60] (see SI on data section).

Another caveat is that funding acknowledgement practices may vary across countries and disciplines[31]. For instance, previous studies argued that publications with Chinese affiliations have higher rate of funding acknowledgement and are associated with higher number of grants[47,48], although the extent of such biases is not yet clear and evidence tends not to be well-established (see Supplementary information on data section). Third, our estimation of influence may be too simplistic; future work may be able to devise more sophisticated causal inference techniques to estimate the extent of influence that one country is exerting on another. Despite these limitations, our systematic examination of global funding landscape with the best available data allows us to map contrasting funding patterns on a global scale and understand how countries are interconnected through funding.

**Acknowledgments**

L.M. and Y.Y.A acknowledge funding support from the Air Force Office of Scientific Research under award number FA9550-19-1-0391. L.M., V.L., Y.Y. A. and C.R.S. acknowledge funding support from the National Science Foundation under award number 2241237. The funder had no role in study design, data collection and analysis, decision to publish or preparation of the manuscript. We thank Staša Milojević, Byungkyu Lee, Tao Zhou, Junming Huang, and Jian Gao for helpful discussion and comments.

**Author Contributions:** L.M. and F.W. conceived the study; all authors contributed to the design of the study; V.L. and F.W. prepared the primary datasets; L.M. and F.W. performed analysis; all authors contributed to the interpretation of the results and writing of the manuscript.

**Competing Interest Statement:** The authors declare no conflict of interest

**Data and materials availability:** Restrictions apply to the availability of the bibliometric data, which is used under license from Thomson Reuters. Readers can contact Thomson Reuters at the following URL: http://thomsonreuters.com/en/products-services/scholarly-scientific-research/scholarly-search-and-discovery/web-of-science.html. The code used for data processing and analysis will be available https://github.com/LiliMiaohub/national-funding.

# Supplementary Information

**Funding acknowledgement data**

Web of Science (WoS) starts to record funding data in August of 2008. Considering the quality and completeness of funding data, we utilize records from 2009 onwards. For the same reason, only publications that are in fields of Natural Sciences, Engineering and Medicine are considered. We only consider the document type "Article", "Note" and "Review". In total, 12,759,130 publications are included in our analysis. Among the 12,759,130 publications, 7,737,510 (60.6%) publications have funding information.

Our analysis builds upon the acknowledgement data within the WoS dataset. However, it is important to note that the acknowledgement data in WoS is not exempt from errors. WoS primarily relies on in-text extraction to collect the funding information from acknowledgements, and it is unclear how the WoS deal with cases where funding acknowledgement is found in other sections of the manuscript, such as footnotes. The accuracy of funding acknowledgements varies across disciplines and research grants[1,2]. Powell found that WoS returned around 80% of all publications supported by NIH grants, whereas PubMed returned 93% of them[2]. Koier and Horling found that WoS incorrectly retrieved acknowledgements for about 24% of research publications supported by two Dutch climate programs[3]. After manually extracting funding data from the full text (including other sections that may include funding acknowledgement) of cancer related publications produced by UK affiliated authors in 2011, Grassano et al. find, among all the sampled publications with funding information, WoS reports funding information for 93% of them[4]. Grassano et al. also report that WoS missed at least one funder in about 11% of records[4]. This result is roughly comparable to the results of Álvarez-Bornstein, who found the rate of missing information from acknowledgement in WoS is quite low. For instance, they found that funding information was entirely lost (neither the funder nor the grant number was collected) in 7.1% of sampled articles and is partially lost (only the funder or the grant number was collected) in 5.8% of sampled articles[5]. Wang and Shapira find that the likelihood of misreporting funding information in WoS is relative low for nanotechnology; only one paper is found to incorrectly index the funding field from funding acknowledgement among the 150 sampled publications[6].

Since the quality of the data plays vital role in our analysis, we systematically evaluated the accuracy of funding information retrieval within the WoS. The analysis is performed with a most recent WoS version, which is slightly different from the version that we used for the analysis. We will discuss the consequence of using different versions of WoS in the following paragraph.

In alignment with our main analysis, our robustness analysis uses journal papers and review articles in the Science Citation Index Expanded (SCIE). We assert that Web of Science (WoS) rarely introduces spurious, nonexistent funding information (referred to as the false positive case) when original articles lack funding information or when such information are not reported by external sources. To substantiate this assertion, we conducted a manual examination of papers associated with funding information in the WoS dataset. The manual examination of a randomly sampled set of 30 articles reveals that all 30 articles indeed acknowledged funding, yielding a 95% confidence interval for the true positive rate of 94.3%±5.6%. Therefore, we posit that the occurrence of false positives is infrequent. Consequently, we focus on the false negatives—funded papers acknowledging financial support within publications but not documented in WoS.

We estimate the frequency of false negatives in WoS and use it to estimate the true funding rates. First, we assess the overall long-term funding trend by examining two specific time points, namely 2009 and 2018. For each time point, from all papers that do not have any funding information in the WoS, we randomly sample 150 papers (300 total). We then conducted a manual verification of the funding information in the sampled publications.

For those from 2009, we identify funding support acknowledgements in 24 out of 150 (16%) of them; for those from 2018, 8 out of 150 (5.3%) contain identifiable funding information. To estimate the number of papers should be reclassified as having funding information, we measure the error rate in recognizing papers without funding information, which is defined as $ER = FN/N$, where $FN$ represents the number of papers should be reclassified as having funding information, and $N$ represents the number of papers classified as lacking funding information in WoS.

We applied bootstrapping to the sampled dataset to estimate the error rate and its confidence intervals. In 2009, the estimated error rate is 16.2% (95% CI: [16.0%, 16.4%]), and in 2018, it is 5.4% (95% CI: [ 5.3%, 5.5%]). According to the most recent version of WoS, out of 1,038,638 papers published in 2009, 51% (531,320) are recognized to include funding information and 49% (507,318 papers) are identified as not having funding information. Considering the estimated error rate, approximately 16.2% of papers identified as without funding information should be reclassified. Incorporating misclassified papers, we anticipate that about 59% (613,525 papers, 95% CI: [612,602, 614,449]) of these papers actually have funding information. For the 1,548,696 papers published in 2018, 68% (1,051,390 papers) already contain funding information. Factoring in the estimated error rate, approximately 70% (1,078,350 papers, 95% CI: [1,077,757, 1,078,944]) of papers in 2018 should be classified as having funding information. Despite a relatively high error rate in identifying papers without funding information in earlier years, the results, after adjusting for misclassifications, still support an increasing trend in the proportion of papers containing funding information within WoS (see Figure 1).

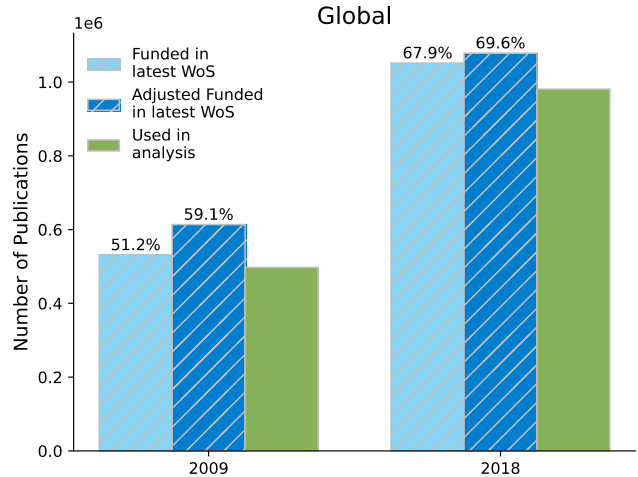

*Figure 1 **The number of publications with funding information**. The light blue bars represent the number of papers identified in the latest version of WoS as containing funding information. The associated percentage represents the corresponding fraction, calculated as $N_{funded}/N_{total}$, where $N_{total}$ is the total number of publications in the latest version of WoS. The dark blue bars represent the number of papers should be classified as have funding information after incorporating false negative cases. The associated percentage represents the corresponding fraction. The green bars represent the number of funded papers in the dataset we used for our analysis.*

We further compared the data coverage in our analysis with the expected number of publications with funding information in the latest WoS release. In the dataset used for the analysis, we identified 497,411 and 980,965 publications with funding information for the years 2009 and 2018, respectively, constituting 81% and 91% of the anticipated number of publications with funding information (see Figure 1). The difference in coverage between our dataset and the most recent WoS arises from three key factors. First, the anticipated number of publications is calibrated to address false negatives, yielding to a more comprehensive coverage. Second, WoS has been updated dynamically, incorporating additional funding information into the dataset. Third, our analysis selected publications based on funding institutions, focusing on funding institutions with occurrences more than two instances. This filtration identified 755,031 articles (6% of all articles in our analysis) as unfunded papers. Although multiple factors have contributed to a more comprehensive coverage of funding information in the latest version of WoS, given that the data used in our analysis encompass substantial amount of funding information compared to the expected number, we believe the data is valid and robust for the analysis.

Given that our analysis primarily focuses on the evolving dominance between the United States and China, we conduct additional estimations to assess the error rates in recognizing papers without funding information for these two countries, defined as $ER_c = FN_c/N_c$ where $FN_c$ represents the number of authored by country c misclassified as lacking funding information by WoS, and $N_c$ represents the number of papers authored by country c classified as papers without funding information by WoS. Sampling approximately 100 papers from China and the United States for the years 2009 and 2018, respectively, we manually cross-validated funding information within these publications. Applying bootstrapping on the sampled dataset, our results indicate an error rate of 29.9% (95% CI: [29.5%, 30.2%]) for Chinese-authored publications and 26.3% (95% CI: [26.1%, 26.6%]) for US-authored publications in 2009. In 2018, these rates are 20.4% (95% CI: [20.2%,20.6%]) for China and 10.7% (95% CI: [10.5%,10.9%]) for the United States.

In the most recent version of WoS, among the 122,394 and 393,720 papers authored by China in 2009 and 2018, 34,350 and 55,973 papers are classified as lacking funding information. Considering the estimated error rate, 26.3% and 10.7% of papers identified as lacking funding information should be reclassified. After incorporating misclassified cases, there are 98,214 and 348,724 Chinese researchers authored papers should have funding information in 2009 and 2018. In our analysis dataset, we have 83,947 and 330,812 papers authored by Chinese authors with funding information in 2009 and 2018, representing 85% and 95% of the expected number of funded publications based on the latest WoS (see Figure 2). For US authors, the latest WoS reports 163,769 and 257,887 papers with funding information in 2009 and 2018. After incorporating misclassified papers, these numbers become 194,704 and 269,189 for US-authored papers. In our analysis dataset, we have 153,889 and 242,645 papers authored by US researchers with funding information in 2009 and 2018, which constitutes 79% and 90% of the expected number of funded papers (see Figure 2).

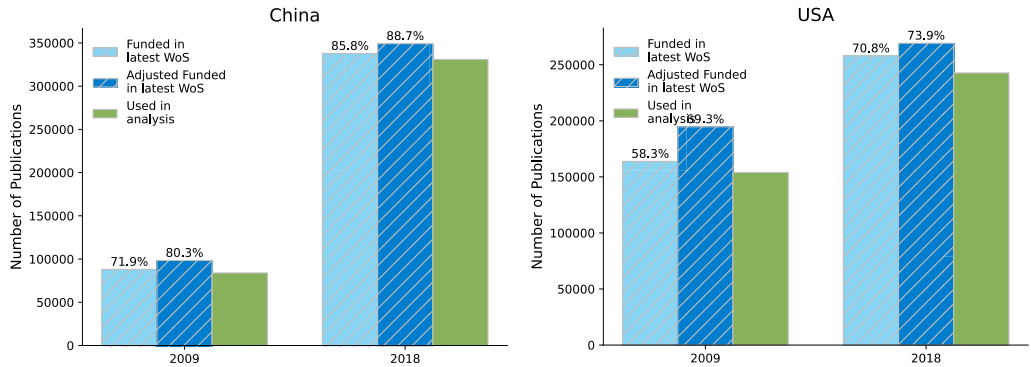

*Figure 2 The number of funded papers by China and the United States. The light blue bars represent the number of papers identified in the latest version of WoS as containing funding information. The associated percentage represents the corresponding fraction, calculated as $N_{funded}/N_{total}$, where $N_{total}$ is the total number of publications authored by the country in the latest version of WoS. The dark blue bars represent the number of papers should have funding information after incorporating false negative cases. The associated percentage represents the corresponding fraction. The green bars represent the number of funded papers in the dataset we used for our analysis.*

To assess the impact of omitted funding data, we estimate the number of publications funded by the United States and China in 2009 and 2018 using the latest WoS data, while incorporating the misclassified cases. However, it is important to note it is infeasible to precisely replicate the calculations from the main analysis (Figure 1d in the main text) due to the involvement of fractionalization of funded publications based on the number of funders and funding instances, where specific funding information of the misclassified publications remain unknown. Therefore, we estimated the funded publications in the latest WoS leveraging the proportion of papers funded by each country, as quantified in our current analysis. Specifically, we computed the proportion of papers funded by country c when researchers from country c are listed as authors, denoted as $P_1 = F_{c,t}/N_{c,t}$ where $F_{c,t}$ represents the number of papers country c funded when researchers from country c are listed as authors (calculated using fractional counting, see Method), and $N_{c,t}$ is the number of funded papers country c authored. Similarly, the proportion of papers funded by country c when researchers from country c are not listed as authors is defined as $P_2 = Q_{c,t}/N_{\neg c,t}$ where $Q_{c,t}$ is the number of papers country c funded when researchers from country c are not listed as authors and $N_{\neg c,t}$ is the number of funded papers that country c is not listed in authorship country. The number of papers funded by country c is estimated as: $M_{c,t} \times P_1 + M_{\neg c,t} \times P_2$ where $M_{c,t}$ and $M_{\neg c,t}$ represent the number of funded papers authored by country c and not authored by country c in the latest WoS after adjusting for false negative rates.

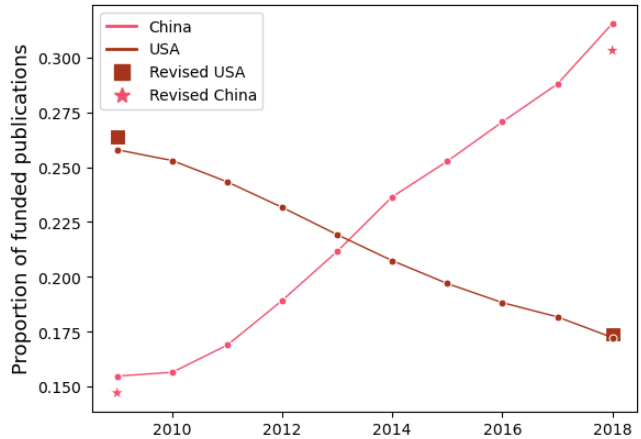

*Figure 3 The proportion of funded publications that are funded by the United States and China. Solid lines depict the trend derived from the dataset used in our analysis, while squares and stars denote the corresponding proportion for the United States and China derived from the latest WoS, accounting for false negative cases.*

Our finding indicates that, despite data issues, our original results are remarkably robust. Moreover, since the dataset employed in our analysis covers a substantial portion of the "ideal funding data", we believe that the funding portfolio of countries, as illustrated in Figure 2 in the main text, remains robust against omitted data. Lastly, considering the high and slightly higher funding coverage for China compared to the United States, the conclusion highlighting the more influential impact of funding from the United States remains unchallenged by the omitted data.

Finally, to comprehensively estimate the quality of funding information in the WoS database, we sampled an additional 500 publications from the WoS dataset and compared the disclosed funding information of these papers with their corresponding entries in the Dimensions data. Within the sampled 500 papers, we find that 131 of them contain funding information from WoS, while only 51 of these papers contain funding information within the Dimensions dataset. There is an overlap of 36 publications that contain funding information in both datasets. A subset of 95 papers is identified as possessing funding information in the WoS while lacking corresponding data in the Dimensions dataset. We examined 10 randomly selected papers from this subset and find each of these 10 papers indeed acknowledged funding support. Meanwhile, our examination also reveals that the WoS dataset is not entirely flawless. We conducted another examination with the 15 articles within the Dimensions dataset that contained funding information but lacked corresponding information in WoS. We find 12 out of the 15 articles contain funding information within the paper. Two articles lack funding information and Dimensions inaccurately identified the funding for one paper. The results collectively suggest that although funding information in the WoS dataset is not exempt from errors, it is a reasonable dataset to investigate the global funding landscape.

**Accuracy of nationality identification**

We applied two methods to evaluate the accuracy of the identification. First, we crosschecked our identification result with other curated datasets. The first data we used is the unified and cleaned list from WoS[1]. The list contains 1,254 funding agencies where 1,119 (89.2%) organizations are included in our analysis. Among the 1119 funding agencies, 1074 (95.98%) organizations are assigned to the same countries as the WoS list. There are five organizations that are assigned to the incorrect country in WoS. In our dataset, 40 agencies have different countries affiliations with the information contained in WoS list. Among the 40 incorrectly identified agencies, 12 (12/40) institutions are identified as either "EU" or "Multi-nation" in our method while they are assigned to the country where the headquarter of the organization locates. There are 8 (8/40) institutions are incorrectly identified through the step one where we extracted the country relevant information from the name of the institution. For example, "American University Cairo" is assigned to the US by us while in actual it is a university locates in Egypt. For the remaining 20 institutions, they all are incorrectly identified using the second step where the authorship institution-level information is used. A potential bias in the second step is the method favors big collaborative countries, particularly when it is dealing with the small research grants with a few publications. For instance, among the 20 incorrectly identified institutions, half of them are incorrectly assigned to the US due to the international collaboration advantage of US institutions.

The second dataset we used is a set of global cancer research funding institutions . This dataset consists of funding institutions that fund cancer research. The list of funding agencies is collected from five different sources include institutions extraction from cancer related google news, bibliometric approaches using WoS, private for-profit financial entities for cancer from the Pharmaceutical Research and Manufacturers Association, funding institutions in the Union for International Cancer Control, and funding organizations from the US Internal Revenue Service. The multiple sources yield 4737 funding agencies in total. Since the cancer funding agency list contains institutions that are not frequently acknowledged by academic publications such as the for-profit financial entities, among the 4737 funding agencies, only 2501 (52.9%) appear in our analysis. Among the 2501 institutions, 2309 (92%) institutions are identified as the same countries. There are 16 European academic associations assigned to the location of the headquarter: e.g., the European Institute of Oncology is assigned to Italy. Instead of assigning a single country, we label all EU associated agencies as "EU". There are 26 institutions that are assigned to the wrong country in the cancer list. For instance, "University of Liverpool" is assigned to the US while it is located in the UK. Therefore, including all the "EU" institutions and the incorrectly identified institutions by the cancer research, a total of 2351 (94%) institutions are correctly identified by our list. For the rest of the 150 incorrectly identified agencies, 19 (12.7%) are identified through the first step where the country information in the name string is used and the rest of them are identified through the second step using the funded author information. Among the 131 mistakenly assigned institutions by the second step, 61 (46.57%) institutions are identified to US agencies which reinforces that the author information identification favors scientific advanced countries due to their collaboration advantage.

To further estimate the bias that is introduced by our identification methods, particularly to the US, we manually validated 100 institutions that are randomly selected from the institutions that are assigned to the US and China, respectively. A further validation shows our identification has high accuracy. Among the 100 institutions we sampled from the US, five of them are incorrectly identified; one is identified with step one and four are identified with step two (where the authorship institution information is used). There are eight funding institutions where we are unable to find relevant information. In total, these 13 institutions funded 71 publications which is only 3% of the publications that are covered by the sampled institutions. Among the 100 institutions we sampled from China, two of them are incorrectly through the step two and 11 of them are unidentified. In total, these 13 institutions cover 76 publications which is only 1% of the publications that are covered by the sample.

---

[1] https://support.clarivate.com/ScientificandAcademicResearch/s/article/Web-of-Science-Core-Collection-Availability-of-funding-data?language=en_US

**Country's research funding intensity by source of the fund**

As showed in figure 6, North American, African and Oceanian countries have the lowest proportion of publications exclusively funded by the focal authorship countries, while concurrently having the highest proportion of publications exclusively funded by foreign countries. We conduct further examination of countries with a significant share of domestic funding within each region. The results reveal that China has the highest proportion of publications funded domestically across all countries. Despite the generally lower percentage of domestically funded publications in North American countries, the United States, Canada and Mexico emerge as notable outliers with high proportion of domestically funded publications (see Table 1).

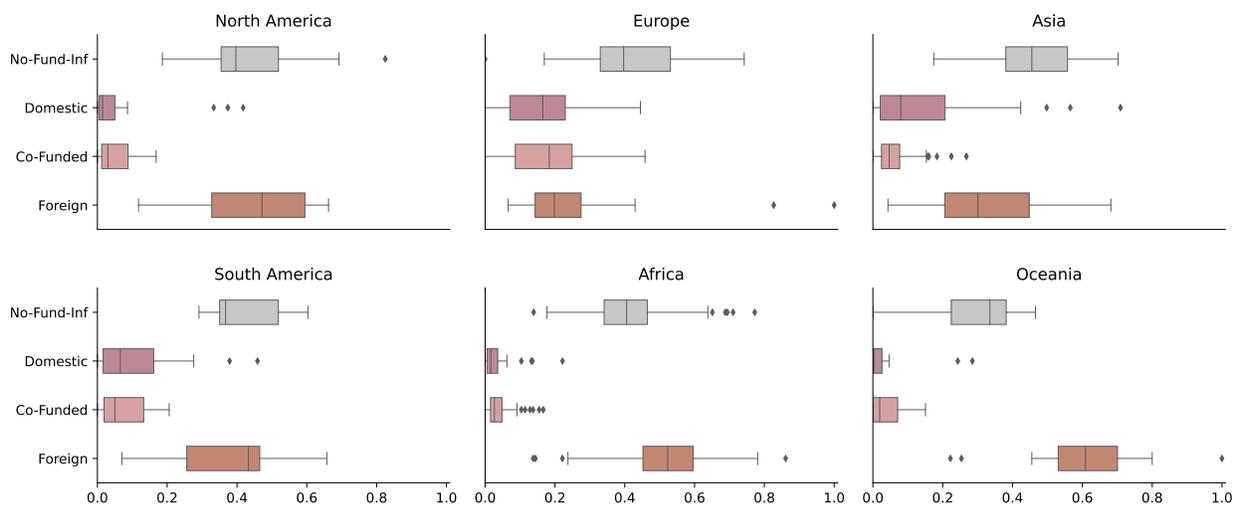

*Figure 4 Distribution of funding portfolios across regions. "No-Fund-Inf" is the abbreviation for "No Funding Information", denoting papers that lack explicit funding information within the WoS dataset. "Domestic" represents publications that are exclusively funded by agencies from the authorship country. "Co-Funded" represents publications that are co-funded by the focal authorship country and other countries. "Foreign" represents publications that are exclusively funded by foreign countries.*

*Table 1 Funding portfolio of outlier countries across regions based on domestic funding*

| cntry | region | Co-Funded | Domestic | Foreign | No-Fund-Inf |
|---|---|---|---|---|---|
| **Ethiopia** | Africa | 0.05 | 0.14 | 0.38 | 0.43 |
| **Mauritius** | Africa | 0.10 | 0.10 | 0.35 | 0.45 |
| **South Africa** | Africa | 0.17 | 0.22 | 0.26 | 0.35 |
| **Tunisia** | Africa | 0.04 | 0.13 | 0.14 | 0.69 |
| **China** | Asia | 0.07 | 0.71 | 0.04 | 0.17 |
| **South Korea** | Asia | 0.08 | 0.57 | 0.07 | 0.29 |
| **Taiwan** | Asia | 0.11 | 0.50 | 0.11 | 0.29 |
| **Canada** | North America | 0.17 | 0.33 | 0.19 | 0.31 |
| **Mexico** | North America | 0.12 | 0.37 | 0.13 | 0.37 |
| **United States** | North America | 0.13 | 0.42 | 0.12 | 0.34 |
| **Australia** | Oceania | 0.15 | 0.28 | 0.22 | 0.34 |
| **New Zealand** | Oceania | 0.12 | 0.24 | 0.25 | 0.38 |
| **Argentina** | South America | 0.17 | 0.38 | 0.16 | 0.29 |
| **Brazil** | South America | 0.10 | 0.46 | 0.07 | 0.37 |

**Estimating a country's dependence on international funding**

To investigate a country's dependence on international research funding, we construct the first counterfactual publication record by removing publications involving any degree of international funding. However, considering the potential situation wherein researchers can leverage domestic funding in the absence of foreign financial support, we build the second counterfactual publication record by removing publications exclusively funded by foreign sources. This additional experiment assesses countries' reliance on international funding under the assumption that domestic funding can sustain research production even in the absence of foreign funding. The results indicate that, if additional support from domestic funding agencies is possible, African and Oceanian countries till bear the most significant impact during a disruption in international funding (see figure 7). This reaffirms the vulnerability of African and Oceania countries to funding disruptions.

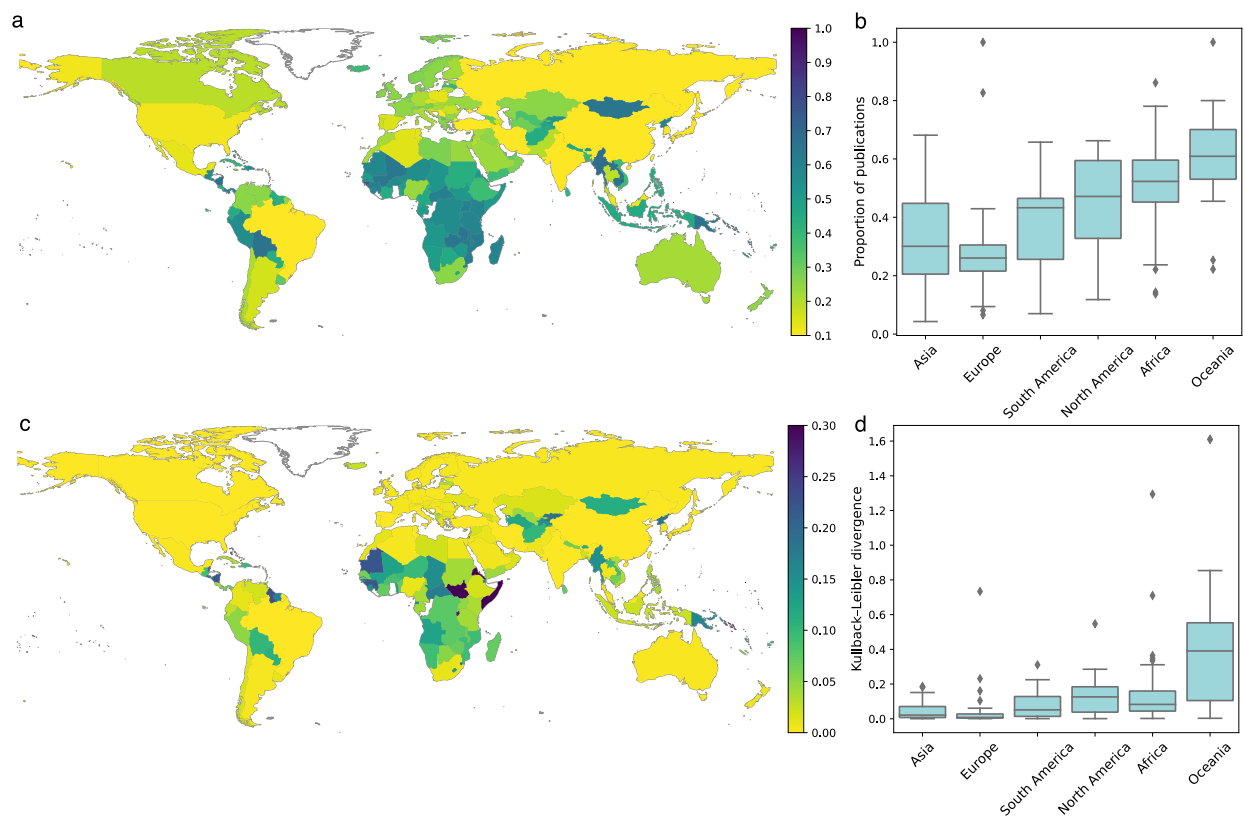

*Figure 5 **The impact of removing publications funded exclusively by internationally funding. Asian countries experience the least lost while African countries as well as Oceania countries suffer the largest lost.** (a) The proportion of reduced publications after the exclusively internationally funded publications are removed. (b) the country-level publication reduction grouped by continents. (c) The difference between actual research profile and the counterfactual research profile. The difference is measured by the Kullback-Leibler divergence. (d) the profile change of countries by continents.*

## Estimating a country's global impact

To investigate a country's dependence on international research funding, considering the possibility that researchers can potentially access funding resources from alternate countries in the absence of financial support from a specific country, we build another counterfactual publication record by removing publications exclusively funded by the focal country. The analysis measures the impact of the focal country under the condition that the focal country is the sole provider of financial resources for the paper's production, with no other funding sources available for research production. The results again demonstrate that removing funding from the United States would have the most significant impact on the scientific production of other countries. In comparison, the impact of China is considerably less substantial (see figure 8). The results reaffirm the conclusion that the United States has been the leading scientific investor in other countries.

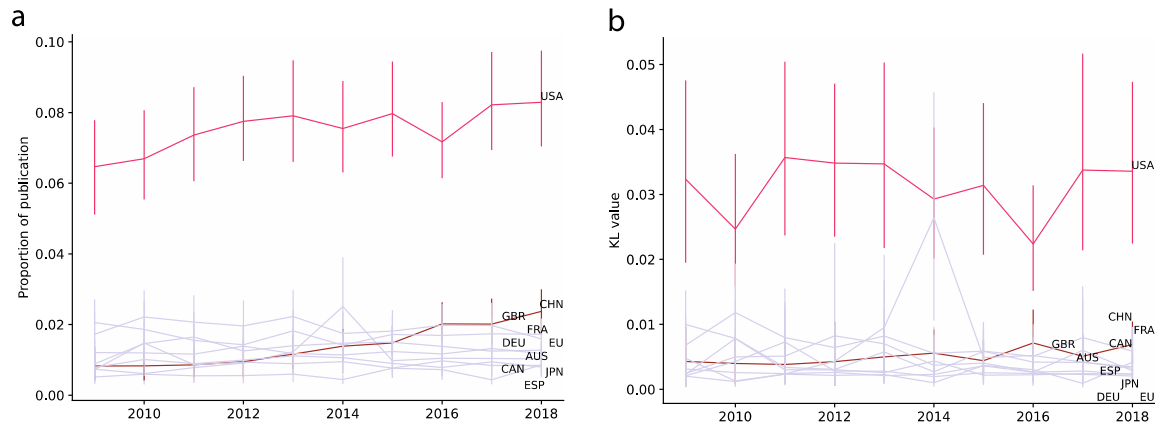

*Figure 6 Estimating a country's impact on global scientific production by removing publications that are exclusively funded by the focal country.*